\documentclass{PoS}

\title{Renormalisation of quark bilinears with
$\mathbf{N_f=}2$ Wilson fermions and tree-level improved gauge action}

\ShortTitle{RCs for $N_f=2$ Wilson fermions and tree-level improved gauge action  
}

\author{\speaker{P. Dimopoulos}~, ~~R. Frezzotti, ~~G. Herdoiza,  
                  ~~A. Vladikas\\
         Dip. di Fisica, Universit{\`a} di Roma Tor Vergata and INFN,
      Sez. di Tor Vergata,\\ Via della Ricerca Scientifica, I-00133 Roma, Italy \\ 
      E-mail: \\ \email{\{dimopoulos,frezzotti,herdoiza,vladikas\}@roma2.infn.it, 
                     }}

\author{V. Lubicz, ~~S. Simula\\
        Dip. di Fisica, Universit{\`a} di Roma Tre and INFN, Sez. di
      Roma Tre,\\ Via della Vasca Navale 84, I-00146 Roma, Italy \\
        E-mail: \email{lubicz@fis.uniroma3.it,simula@roma3.infn.it}}

\author{M. Papinutto\\
        CERN, Physics Department, Theory Division,CH-1211 Geneva 23, Switzerland \\
        E-mail: \email{mauro.papinutto@cern.ch}}

\author{On behalf of the ETM Collaboration} 

\abstract{We present results for the renormalisation constants of bilinear quark operators,
using the $N_f=2$ twisted mass 
Wilson action at maximal twist (which guarantees automatic $O(a)$ improvement) 
and the tree-level Symanzik improved gauge action.
The scale-independent renormalisation constants are computed with a new method, 
which makes use of both standard twisted mass and Osterwalder-Seiler fermions. 
Moreover, the results from an RI-MOM calculation are presented for both scale 
independent and scale dependent renormalisation constants. 
          }

\FullConference{The XXV International Symposium on Lattice Field Theory\\
		 July 30-4 August 2007\\
		 Regensburg, Germany}

\usepackage{subfigure}
\begin{document}

\section{Introduction}
We compute quark bilinear renormalisation constants (RCs),
based on the ETMC $N_f=2$ dynamical quark action which consists of a tree--level 
improved Symanzik gauge action and twisted mass (tm) Wilson fermions at maximal 
twist \cite{etmc_letter}.  Our results are
automatically improved in the spirit of ref.~\cite{FR}.  In section 2 we present a new
(non-perturbative) method for the calculation of the scale independent RCs, $Z_A$ and $Z_P/Z_S$,
based on the use of two valence quark actions and a standard calculation of $Z_V$ within 
the tm valence quark sector. In section 3 we describe the RI-MOM
calculation of all RCs (both scale dependent and scale independent ones).

\section{Calculation of the scale independent RC}

In this section we present a calculation of the scale independent RCs, namely $Z_V$,
$Z_A$, $Z_P/Z_S$. The evaluation of $Z_V$  is based on the PCAC Ward identity method
(see refs.~\cite{ZV_calc} for details). This calculation leads to very precise results. The
computational method for $Z_A$ and $Z_P/Z_S$ is new. It is based on the use of 
two regularisations for the valence quark actions. 
One is the standard twisted mass action,
while the other is the Osterwalder--Seiler (OS) variant \cite{OS}. In the so called physical basis
these actions  can be compactly  written in the form:
\begin{equation} \label{action}
S_{val} = a^4 \sum_{x} \bar\psi(x) (\gamma \tilde{\nabla} -i \gamma_5 ~r~ W_{cr} + \mu_q ) \psi(x) \,\,\, ,
\end{equation}
with $ W_{cr} = -\frac{a}{2} \sum_{\mu} \nabla_{\mu}^{*} \nabla_{\mu} + M_{cr}(r=1)$,
$\psi = (u ~~d)^{T}$,  $r = {\rm diag}(r_u ~~r_d)$ and $\mu_q={\rm diag}(\mu_u ~~\mu_d)$.  
The twisted mass case  corresponds to $r_u = -r_d = \pm 1$, while the Osterwalder-Seiler  
case is obtained taking $r_u =  r_d = \pm 1$. Sea quarks are regularized in the standard tm framework.

Consider that, for the two different choises of the matrix $r$, we perform 
the following two  axial  transformations of the quark fields, namely 
$(u,d)=\exp[i( \gamma_5 \tau_3 \pi/4)] (u^{'},d^{'})$ and $(u,d)=\exp[i( \gamma_5 \pi/4)] (u^{'},d^{'})$,
respectively.
Each of  the  actions (\ref{action}) transforms respectively  into an action with the Wilson term 
in the standard form (no $\gamma_5$ and no $\tau_3$). This is a rotation into the tm basis at maximal twist. 
However the tm action has a mass term of the form $i\mu \bar{\psi^{'}}\gamma_5 \tau_3 \psi^{'}$, while
the OS one has $i\mu \bar{\psi^{'}}\gamma_5  \psi^{'}$.
Consider, now,  an operator $O_{\Gamma}$ defined in the physical basis. Under the  two 
axial trasformations this operator transforms  into two operators, called 
$O_{\tilde{\Gamma}}$ and $O_{\tilde{\tilde{\Gamma}}}$, which, in general,  are not of the same form. 
However the respective renormalised matrix elements between given physical states have to be equal up to $O(a^2)$ effects. 
This is due to the fact that  in the continuum limit each of them should coincide, up to $O(a^2)$, 
with the corresponding matrix element of the unique physical operator, $O_{\Gamma}$. Therefore, if we call 
$Z_{O_{\tilde{\Gamma}}}$ and $Z_{O_{\tilde{\tilde{\Gamma}}}}$ the respective 
renormalisation constants for the two operators,  
we  have:
\begin{equation}\label{renorm}
 Z_{O_{\tilde{\Gamma}}} \langle O_{\tilde{\Gamma}} \rangle^{tm}  = 
 Z_{O_{\tilde{\tilde{\Gamma}}}} \langle O_{\tilde{\tilde{\Gamma}}}\rangle^{OS} + O(a^2) ~~.
\end{equation}     
Renormalisation constants are named, as usual, after the basis in which the Wilson term has
its standard form.
For maximal twist, the operator renormalization pattern in the physical and twisted bases 
is shown in Table 1 for both OS and tm formalisms. The primed operators refer to the tm basis 
while the
unprimed ones to the physical basis and we have adopted the notation,  
$O_{\Gamma}=\bar{u}\Gamma d$, for both the primed and unprimed case.

\begin{table}
\begin{center}
\begin{tabular}{ccccccccccc}
\hline \hline
               OS case &&&&&&&&&&                tm case    \\
$(A_R)_{\mu, ud} = Z_A A_{\mu, ud} =  Z_A A_{\mu, ud}^{'}$  
&&&&&&&&&& $~~~(A_R)_{\mu, ud} =  Z_V A_{\mu, ud} =  -i Z_V V_{\mu, ud}^{'}$ \\
$(V_R)_{\mu, ud}  = Z_V V_{\mu, ud} =  Z_V V_{\mu, ud}^{'}$  &&&&&&&&&&
$~~~~(V_R)_{\mu, ud} = Z_A V_{\mu, ud} =  -i Z_A A_{\mu, ud}^{'}$ \\
$(P_R)_{ud}      = Z_S P_{ud}      ~=  i Z_S S_{ud}^{'}$     &&&&&&&&&&
$(P_R)_{ud}     = Z_P P_{ud}     =  Z_P P_{ud}^{'}$         \\
\hline \hline
\end{tabular}
\end{center}
\caption{Renormalization pattern of the bilinear quark operators for the OS and tm case at 
maximal twist.}
\label{table_dict}
\end{table}

{\bf Calculation of $Z_P/Z_S$:}
Our method is based on comparing the  amplitude  $g_{\pi} = <0|P|\pi>$,
computed both in tm and OS formalisms. We start by considering, 
in the physical basis, the correlator 
$C_{PP}(t) \equiv \sum_{\bf x} <\bar{u}\gamma_5 d (x) ~ \bar{d}\gamma_5 u (0)>$, 
which at large times behaves like
$ C_{PP}(t) \simeq \frac{|g_{\pi}|^2}{2m_{\pi}} [\exp(-m_{\pi} t) + \exp(-m_{\pi} (T-t)) ]$.
In the twisted basis, this corresponds to $C_{S^{'}S^{'}}(t)$ in the OS case and
$C_{P^{'}P^{'}}(t)$ in the tm one. Based on Table \ref{table_dict},  this translates into
\begin{equation} \label{g}
     [g_{\pi^{\pm}}]^{cont} \,\,   = \,\, Z_P \, [g_{\pi^{\pm}}^\prime]^{tm} + O(a^2) 
     \,\, = \,\, Z_S \, [g_{\pi}^\prime]^{OS} + O(a^2) \,\,\,\, ,
\end{equation}
from which the ratio $Z_P/Z_S$ is extracted. 

{\bf Calculation of $Z_A$:}
We undertake the calculation of   $f_{\pi}$ in both OS and tm regularisations. 
In the tm case we use the Ward identity evaluation of the
decay constant: $f_{\pi^{\pm}}^{tm} = 2 \mu_q g_\pi/m_{\pi}^2$.
Note that in this case no renormalisation constant is needed \cite{frezzotti}.
Thus the pion decay constant can be extracted from the large time asymptotic behaviour
of $C_{PP}(t)$ as it is discussed above.

For the OS case we use the correlators $C_{PP}$ and $C_{A_0P}$ (with $r_u=r_d=\pm1$).
The large time asymptotic behaviour of the former correlator has been discussed above,
while the latter goes like
$C_{A_0P}(t) \simeq \frac{\xi_{A_0P}}{2m_{\pi}} [\exp(-m_{\pi} t) - \exp(-m_{\pi} (T-t)) ]$.
Combining these, we can extract the bare OS estimate of the pion decay constant as
$f_{\pi}^{OS}  = \xi_{A_0P} / g_{\pi} m_{\pi}$. Since the tm and OS determinations
of the (properly normalized) decay constant satisfy the relation
\begin{equation} \label{fp}
[f_{\pi^{\pm}}]^{cont} \,\, = \,\, f_{\pi^{\pm}}^{tm} +O(a^2) \,\, = \,\, Z_A f_{\pi}^{OS} + O(a^2) \,\,\, ,
\end{equation}
an estimate of $Z_A$ is readily obtained. Since all computations are performed at finite mass, the final results
for $Z_A$ and $Z_P/Z_S$ are finally obtained by extrapolation to the chiral limit. Moreover,
maximal twist ensures that cut--off effects are of order $O(a^2)$ (\cite{FR},\cite{OS}).

\subsection{Results}
\label{res1}

\begin{figure}[!h]
\begin{center}
\subfigure[]{\includegraphics[scale=0.24,angle=-90]{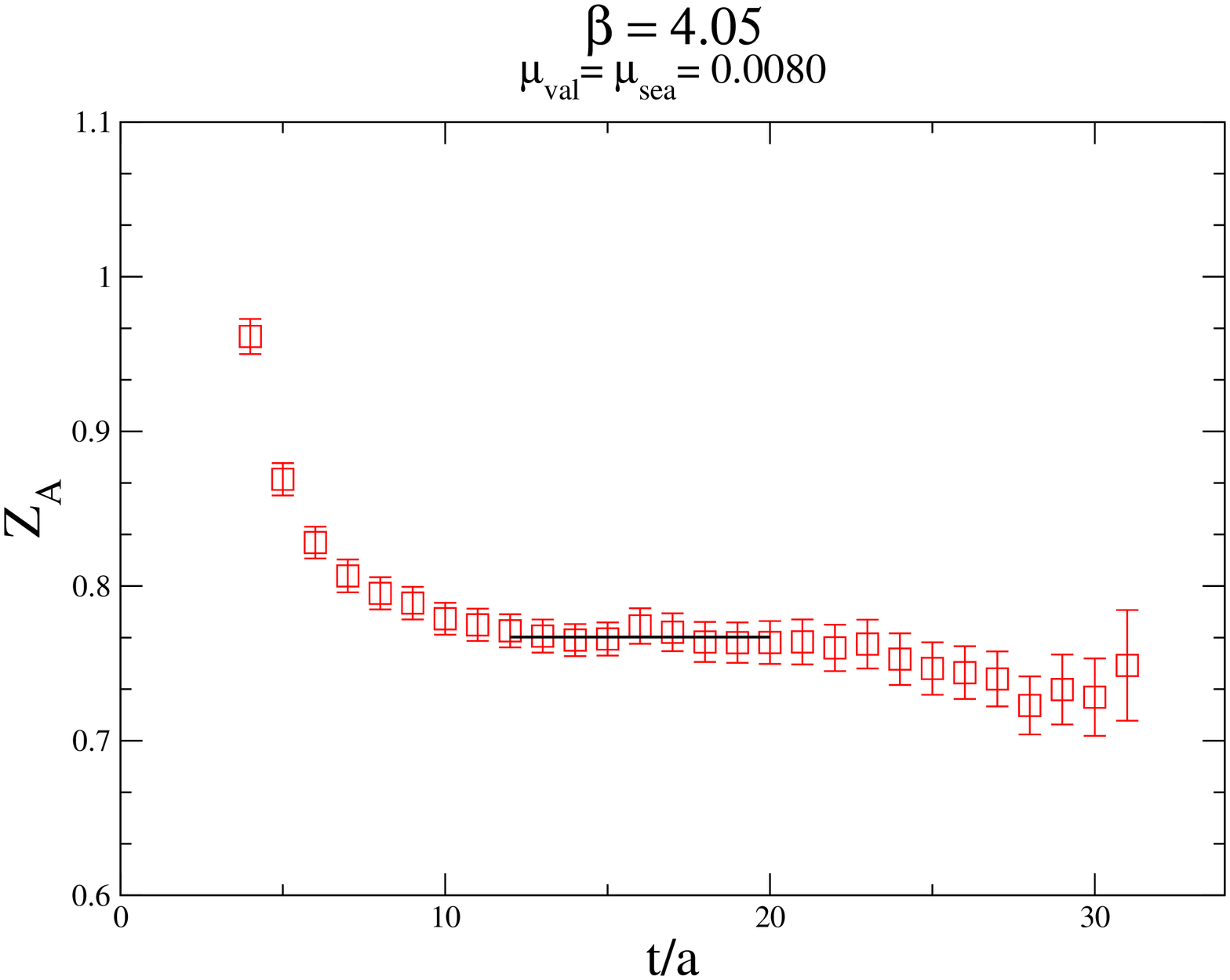}}
\subfigure[]{\includegraphics[scale=0.24,angle=-90]{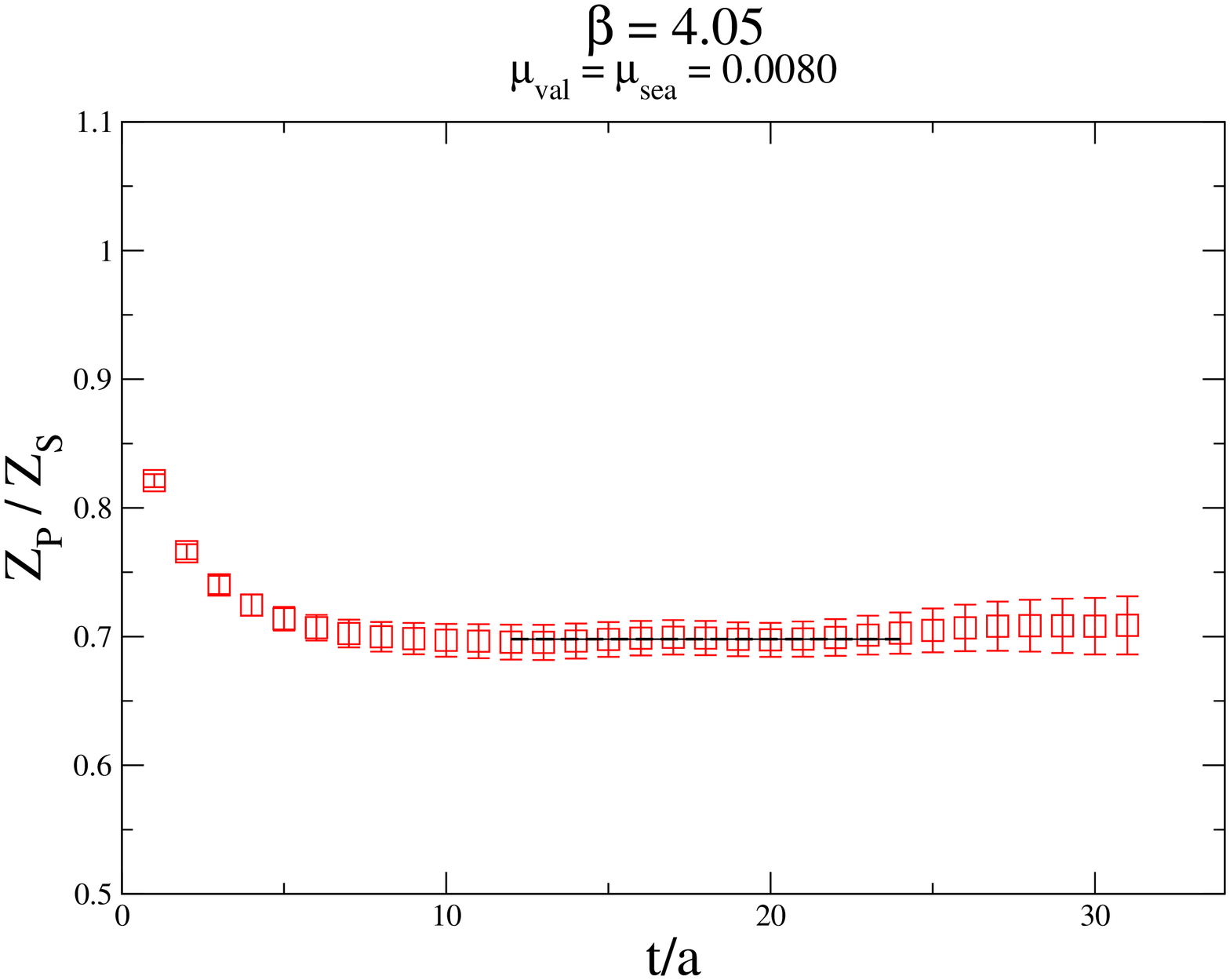}}
\caption[]{Asymptotic behaviour of scale independent normalisation constants}
\label{fig_Ztov}
\end{center}
\end{figure}

Our configuration ensembles for $N_f=2$ sea quarks have been generated at three values of the 
gauge coupling, $\beta=3.80,~ 3.90 ~\mbox{and}~ 4.05$, corresponding to  
lattice spacings $a \sim 0.10, 0.09$ and 0.07 fm. We have performed
240 measurements for the two smallest $\beta$-values and 150 measurements for the
highest one. 
In order to significantly reduce autocorrelation times, correlators were computed every 
20 trajectories (each having trajectory length equal to
$\tau =1/2$). Five sea quark masses have
been simulated at $\beta=3.90$ and four at the other two couplings. The smallest 
sea quark mass corresponds to a pion of about 300 MeV and the 
higher one is just above half the strange quark mass. Eight valence quark masses were used at each
coupling; the lowest ones are equal to the sea quark masses, whereas the others rise to
the region of the strange quark mass. For the inversions in the valence sector 
we have made use of the stochastic method (one--end trick of ref.~\cite{chris}) in order to
increase the statistical information.
Propagator sources are at randomly located timeslices. This turned out to be an optimal way to reduce the 
autocorrelation time. Typical plots on the quality of the signal for the RCs 
(for fixed values of the bare coupling and masses)
are shown in Fig.~\ref{fig_Ztov}. 

Three methods were implemented in the RC computation.
The first consists in calculating the RCs at fixed value of the sea quark mass 
for a number  of valence quark masses and taking the ``valence chiral limit"\footnote{First
and second degree polynomial fits in $\mu_{val}$ have been performed. }. 
Subsequently, the RCs were quadratically extrapolated to the sea quark chiral limit
\footnote{A quadratic dependence on $a\mu_{sea}$ is expected from 
the form of the sea quark determinant, assuming that  lattice artifacts 
on the RCs are not sensitive to  spontaneous chiral symmetry breaking.
However we have verified  that a linear fit in $\mu_{sea}$ leads to
compatible results.}.
The second method consists in inverting the order of the two chiral limits.
The third method is simply the extraction of the RCs from the subset of data satisfying $\mu_{val}=\mu_{sea}$, 
which allows to reach the chiral limit with one single extrapolation in the quark mass. 
Our results from all three methods
are compatible within one standard deviation. 
\begin{table}
\begin{center}
\begin{tabular}{cccc}
\hline \hline
 $\beta$  &    3.80     &  3.90       & 4.05 \\
$Z_A$     &   0.72(2)(1)   &  0.76(1)(1)    & 0.76(1)(1)   \\
$Z_P/Z_S$ &   0.47(2)(1)   &  0.61(1)(1)    & 0.66(1)(1)   \\
$Z_V$     &   0.5814(2)(2) &  0.6104(2)(3)  & 0.6451(2)(3) \\
\hline \hline
\end{tabular}
\end{center}
\caption{The results for the scale independent RCs for  three values of the
gauge coupling.}
\label{table_all}
\end{table}
We present  preliminary data from the second method, which has fits of better quality, 
in Table \ref{table_all}. The first error is statistical while the second is systematic 
coming from the difference between the central values of the various methods. 
A final analysis will be presented in a forthcoming publication.

\section{RI-MOM calculation}

The RI-MOM method is a non--perturbative, mass independent, renormalisation scheme 
proposed in ref.~\cite{rimom}. For a detailed presentation of various technical aspects see
ref.~\cite{rimom2}.  In our case the scheme consists in fixing the Landau gauge and computing 
the momentum space Green function
\begin{equation} \label{G}
G_{\Gamma}^{ud}(p,p') = \sum_{x,y} \langle u(x) (\bar{u}\Gamma d)_0 \bar{d}(y) \rangle e^{-ip \cdot x + i p' \cdot y} \,\,\, ,
\end{equation} 
for a general quark bilinear operator  $\bar{u}\Gamma d$ (with $\Gamma=A, V, S, P, T$) and the  
propagator is written as, \\
$S_q = \sum_{x} \langle q(x) \bar{q}(0) \rangle e^{-ip \cdot  x}$ with $q=u,d.$
\\ Then the forward amputated Green function,
$\Lambda_{\Gamma}^{ud} = S_u(p)^{-1} G_{\Gamma}^{ud}(p,p) S_d(p)^{-1}~,$
is projected by a suitable projector $P_{\Gamma}$ (essentially a properly normalized Dirac matrix).
The RCs, $Z_\Gamma$ and $Z_q$ are obtained by imposing the RI-MOM renormalization conditions
\begin{equation} \label{rimom_cond}
Z_{\Gamma}^{ud} (Z_{u}Z_{d})^{-1/2} \Gamma_{\Gamma}^{ud}(p)|_{p^2=\mu^2} \equiv Z_{\Gamma}^{ud} (Z_{u}Z_{d})^{-1/2}Tr[\Lambda_{\Gamma}^{ud}
P_{\Gamma}]|_{p^2=\mu^2} = 1,~~  
Z_q \frac{i}{12} Tr \left[ \frac{\displaystyle{\not} p S_{q}(p)^{-1}}{p^2} \right]_{p^2=\mu^2} = 1.
\end{equation}
The computation is done for fixed quark masses. The results are extrapolated
to the chiral limit. The renormalisation scale $\mu$ has to satisfy the condition: 
 $\Lambda_{QCD} \ll \mu \ll \pi/a$.

The RCs, calculated in the chiral limit in the way described above,  
are $O(a)$ improved at large momenta \cite{rimom2}.
Moreover an analysis based on the symmetries of MtmLQCD and the O(4) 
symmetry of the underlying continuum theory shows that
$\Gamma_\Gamma^{ud}(p)$ and $\Gamma_\Gamma^{du}(p)$ are separately
$O(a)$ improved for all momenta. In order to increase
the statistical information, we computed the following combinations:
$Z_{\Gamma}= (Z_{\Gamma}^{ud} + Z_{\Gamma}^{du})/2 ~~~\mbox{and}~~~ Z_{q} = (Z_{u} + Z_{d})/2~.$

The scale dependent RCs ($Z_P, Z_S ~\mbox{and}~ Z_T$) are obtained
at a reference scale $\mu_0 = a^{-1}$,
by cancelling the scale dependence $\mu$, at a sufficiently high order in perturbation theory: 
\begin{equation} \label{ZG}
Z_{\Gamma}(a\mu_0) = (~Z_{\Gamma}(a\mu) / C_{\Gamma}(\mu)~) ~C_{\Gamma}(\mu_0) ~.
\end{equation}
Here $C_{\Gamma} = \exp{\int^{\alpha(\mu)}~ d \alpha \,\,[ \gamma_{\Gamma} (\alpha) / \beta(\alpha)}]$ and 
$\gamma_{\Gamma}, ~\beta$ are the anomalous dimension of the operator and the beta function
respectively. They are known at N$^2$LO  for $Z_T$ and N$^3$LO 
for $Z_S$ and $Z_P$ \cite{gracey-chetyrkin}.   
     
It is known that the RI-MOM estimate of $Z_P$ is contaminated by the presence of a Goldstone
pole \cite{PGB}. In the twisted mass theory this problem also arises  for $Z_S$, though 
$O(a^2)$ suppressed. All these contaminations are removed in the subtracted Green function \cite{GV}:  
\begin{equation}\label{sub}
\Gamma_{P,S}^{\mbox{sub}}(p^2, \mu_{q_1}, \mu_{q_2}) = 
\frac{\mu_{q_1}\Gamma_{P,S}(p^2,\mu_{q_1}) - \mu_{q_2}\Gamma_{P,S}(p^2,\mu_{q_2})}{\mu_{q_1}-\mu_{q_2}} 
\end{equation}
where $\mu_{q_1}, \mu_{q_2}$ are non--degenerate valence quark masses.

\subsection{Results}

The simulation parameters are the same as those of  section \ref{res1}. The RCs, computed at fixed
sea quark mass and several valence quark masses, are first linearly extrapolated to the valence
chiral limit. Subsequently, the sea quark chiral limit is obtained by linear extrapolation in $\mu_{sea}^2$.
\begin{figure}[!h]
\begin{center}
\subfigure[]{\includegraphics[scale=0.25,angle=-90]{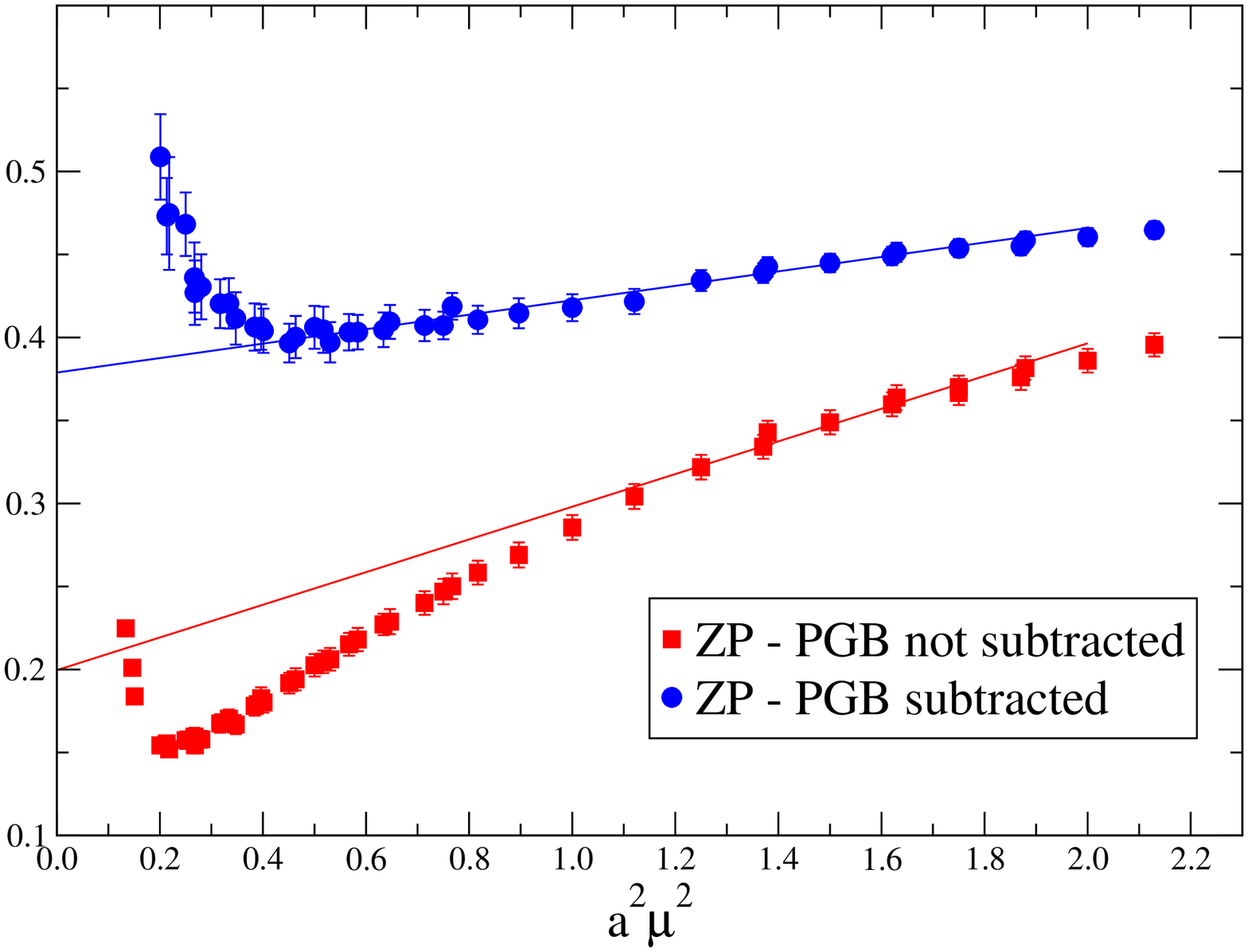}}
\subfigure[]{\includegraphics[scale=0.25,angle=-90]{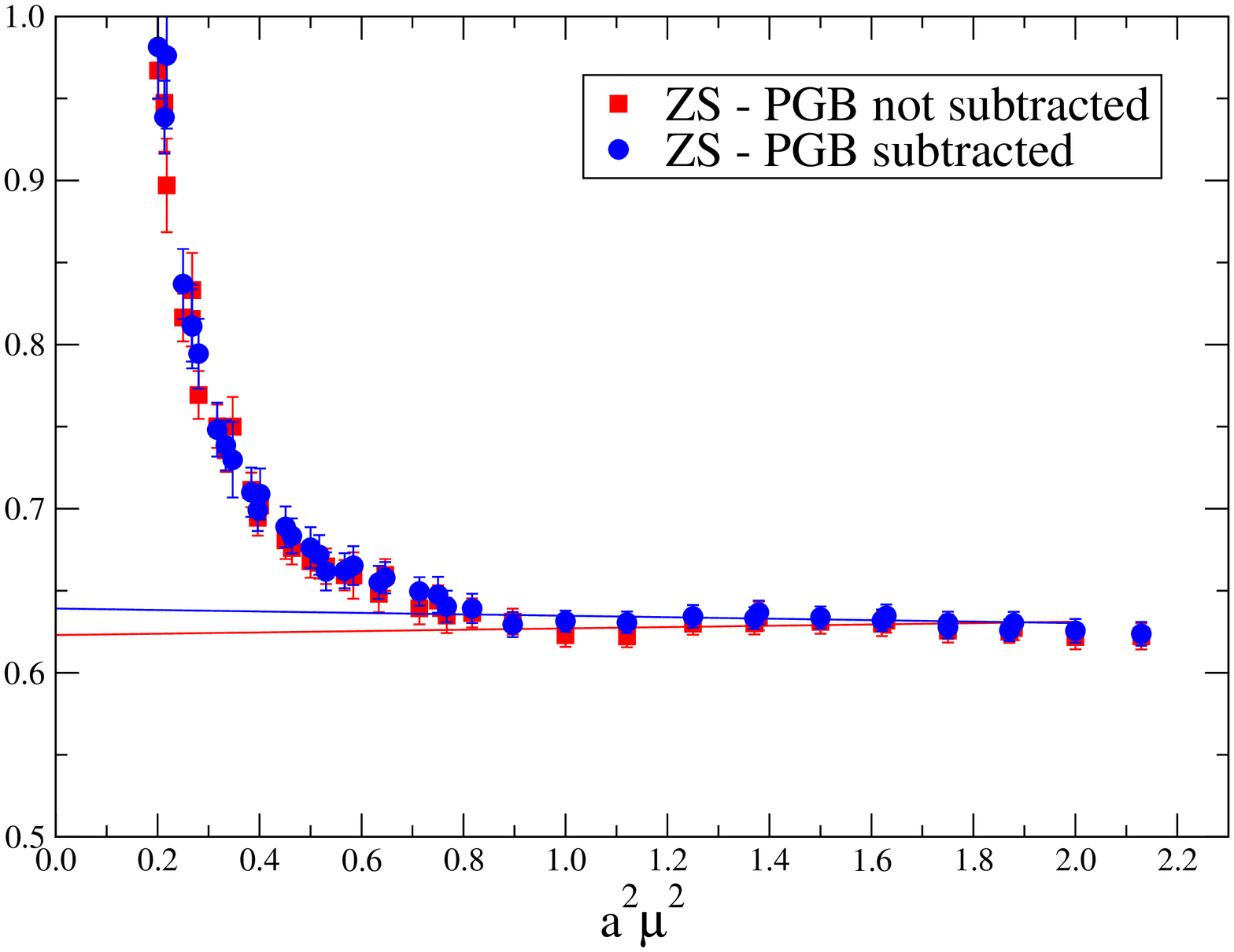}}
\caption[]{Goldstone pole subtraction at $\beta=3.90$; $Z_{\Gamma}(a\mu_0)$ of 
Eq. (\ref{ZG}) is plotted against $(a\mu)^2$.}
\label{fig_sub}
\end{center}
\end{figure}
In  Fig.~\ref{fig_sub} we show the effect of the Goldstone boson subtraction for $Z_P$ and $Z_S$  
for which the  subtracted Green function of Eq.~(\ref{sub}) has been used; 
we see that this has an important effect on $Z_P$, while $Z_S$ is almost unaffected, 
as expected. 
Moreover, we note from Fig.~\ref{fig_sub}  that once
the scale evolution has been perturbatively divided out, the scalar RC is indeed scale independent,
while the pseudoscalar one is still subject to large discretization effects. These are removed by linear
extrapolation, giving a $Z_P$ final estimate as the intercept of the fit.

In Table \ref{rimom_results} we show our preliminary results for the RCs; for $\beta=3.80$ and 4.05 
the results correspond to the lighter value of the sea quark mass only. For $\beta=3.90$ the results come from
a full analysis in the valence and the sea sector. The first error is statistical and the second is 
systematic due to an estimate of the $O(a)$-contribution to the  quark propagator which induces an $O(a^2)$
correction in the determination of the RCs. 
A better estimate of the systematic errors will be available once we finalize the analysis on all the 
three values of lattice spacing.

A first comparison for $\beta=3.90$ between the results of Tables \ref{table_all} and \ref{rimom_results} shows that
the values of $Z_A$ and $Z_p/Z_S$ are in nice agreement and of comparable statistical 
accuracy. The corresponding $Z_V$ results, 
though compatible within the quoted errors,  show that 
the PCAC Ward Identity estimate is statistically more precise\footnote{A slightly 
different determination based on the same WI taken between two one--pion states gives 
very similar results; for example, for $\beta=3.90$ it is found, $Z_V=0.6109(2)$ 
\cite{simula}. Moreover from the Table 1 we find that 
the value of the ratio $(Z_A/Z_V)^2|_{\beta=4.05}$ is consistent with the one found in \cite{michael}.}.

\begin{table}
\begin{center}
\begin{tabular}{ccccccc}
\hline \hline
    $\beta$  &  $Z_A$        &     $Z_P/Z_S$   &  $Z_V$        &   $Z_P$       &    $Z_S$        &   $Z_T$  \\ 
3.80         &  0.75(3)      &      0.47(3)    &   0.62(4)     &    0.30(1)    &    0.64(2)      &   0.73(4) \\
3.90         &  0.76(2)(1)   &      0.63(2)(3) &   0.65(2)(3)  &    0.39(1)(2) &    0.62(1)(5)   &   0.75(1)(2)\\
4.05         &  0.77(1)      &      0.65(2)    &   0.67(1)     &    0.40(1)    &    0.61(1)      &   0.79(1)  \\
\hline \hline
\end{tabular}
\end{center}
\caption{RI-MOM results for the RCs. 
$Z_P$, $Z_S$ and $Z_T$ are calculated at scale $\mu_0=a^{-1}$ (see Eq. (3.3)).
The results at $\beta=3.80$ and
4.05 are preliminary and the quoted errors are purely statistical in
these cases.}
\label{rimom_results}
\end{table} 
We would like to note that combining the result of $Z_P/Z_S$ from the first method with that of $Z_S$
from RI-MOM, an alternative evaluation of $Z_P$ can be  obtained, in which
the problem of the pseudoscalar Goldstone boson pole subtraction is
avoided\footnote{We note in passing that  $Z_P^{-1} = Z_{\mu}$ is
the quark mass renormalisation constant in the twisted mass theory.}. 
For example, for $\beta=3.90$ this calculation gives
$Z_P=0.38$ which is compatible, within the errors,  with the  corresponding value 
given by the  RI-MOM calculation (see Table 3). The results of 
a precise statistical analysis will be given in a forthcoming publication.

\noindent {\bf Acknowledgements} \\ 
We wish to thank G.C. Rossi for fruitful discussions and comments on the manuscript. We 
also acknowledge C. Michael for helpful discussions. This work was partially supported by the
EU Contract MRTN-CT-2006-035482 "FLAVIAnet".

\end{document}